\documentstyle[12pt]{article}
\newcommand{\bea}{\begin{eqnarray}}
\newcommand{\eea}{\end{eqnarray}}
\newcommand{\be}{\begin{equation}}
\newcommand{\ee}{\end{equation}}
\newcommand{\vs}[1]{\vspace{#1 mm}}

\renewcommand{\a}{\alpha}
\renewcommand{\b}{\beta}
\renewcommand{\c}{\gamma}
\renewcommand{\d}{\delta}

\newcommand{\dsl}{\pa \kern-0.5em /}

\newcommand{\pa}{\partial}

\newcommand{\nn}{\nonumber\\}

\begin{document}
\topmargin 0pt
\oddsidemargin 0mm



\begin{flushright}

USTC-ICTS-06-03\\


hep-th/0604048\\


\end{flushright}
\vspace{4mm}

\begin{center}

{\Large \bf

Non-SUSY $p$-branes, bubbles and tubular branes }

\vs{8}

{\large J. X. Lu$^a$\footnote{E-mail: jxlu@ustc.edu.cn} and
Shibaji Roy$^b$\footnote{E-mail: shibaji.roy@saha.ac.in}}

 \vspace{4mm}

{\em

$^a$ Interdisciplinary Center for Theoretical Study\\
University of Science and Technology of China, Hefei, Anhui
230026, China\\

and\\

Interdisciplinary Center of Theoretical Studies\\
Chinese Academy of Sciences, Beijing 100080, China

\vspace{3mm}

$^b$ Saha Institute of Nuclear Physics,

 1/AF Bidhannagar, Calcutta-700 064, India}

\end{center}

\vspace{6mm}

\begin{abstract}
\begin{small}
We consider non-supersymmetric $p$-brane solutions of type II
string theories characterized by three parameters. When the charge
parameter vanishes and one of the other two takes a specific value, the
corresponding chargeless solutions can be regular and describe
``bubbles'' in static (unstable) equilibrium when lifted to $d =
11$. In appropriate coordinates, they represent D6 branes with a
tubular topology R$^{1,p}$ $\times$ S$^{6-p}$ when reduced to $d=10$,
called the tubular D6 branes, held in static equilibrium by a
fixed magnetic flux (fluxbrane). Moreover, a `rotation parameter'
can be introduced to either of the above two eleven dimensional
configurations, giving rise to a generalized configuration
labelling by the parameter. As such, it brings out the relations
among non-supersymmetric $p$-branes, bubbles and tubular D6
branes. Given our understanding on tubular D6 branes, we are able to
reinforce the interpretation of the chargeless non-supersymmetric
$p$-branes as representing $p$-brane-anti$p$-brane (or non-BPS
$p$-brane) systems, and understand the static nature and various
singularities of these systems in a classical supergravity
approximation.
\end{small}
\end{abstract}
\newpage
\renewcommand{\a}{\alpha}
\renewcommand{\b}{\beta}
\renewcommand{\c}{\gamma}
\renewcommand{\d}{\delta}
\topmargin 0pt
\oddsidemargin 0mm

String theory contains in its spectrum not only the BPS states and
branes \cite{Horowitz:1991cd,Duff:1994an} but also their non-BPS
counterparts \cite{Sen:1999mg}. In addition, there are unstable
brane-antibrane systems. For the BPS case, both the open string
and the closed string descriptions of these objects are
well-understood \cite{Aharony:1999ti} and they played
complementary roles in the microscopic calculation of black hole
entropy \cite{Strominger:1996sh} as well as the AdS/CFT
correspondence
\cite{Maldacena:1997re,Witten:1998qj,Gubser:1998bc}. However, the
situation is not so satisfactory for the case of non-BPS states
and branes as well as for the brane-antibrane systems. In this
case, although the open string description is fairly
well-understood \cite{Sen:1999mg,Sen:1999nx,Berkovits:2000hf} in
the weakly-coupled region, the corresponding supergravity or
closed string description does not have the same status as in the
BPS case. Some progress in this approach have been made in
refs.\cite{Brax:2000cf,Lu:2004dp}. The hope is that at least some
properties of the non-BPS states and branes as well as the
brane-antibrane systems can be similarly studied once  we have a
satisfactory supergravity description.

The static, non-supersymmetric $p$-brane solutions of
supergravities, i.e., the low energy string theories, in arbitrary
dimensions were constructed in \cite{Zhou:1999nm,Lu:2004ms}. These
solutions have the isometry ISO($p,1$) $\times$ SO($9-p$) and are
characterized by three parameters. When we take the dimensionality
of spacetime to be ten, the corresponding solutions have been
interpreted in \cite{Brax:2000cf} as representing either the
non-BPS D$p$ branes or D$p$ brane-antiD$p$ brane system depending
on the value of $p$ and the type of theory (IIA or IIB) under
consideration. The three parameters for each solution are related
to the mass, charge and presumably also the tachyon  parameter for
the corresponding unstable configuration. Although their precise
relationships to the tachyon parameter are largely unknown, there
are some interesting attempts to relate them to the physical
microscopic parameters of the brane-antibrane systems
\cite{Brax:2000cf,Lu:2004dp,Bai:2005jr,Kobayashi:2004ay}. Various
properties of these solutions as well as their delocalized
versions have been studied in
\cite{Lu:2004xi,Lu:2005ju,Lu:2005jc}.

In spite of many consistency checks performed so far, questions
regarding the static nature and the singularities of these
solutions may still cast doubt on their interpretation as
representing unstable systems such as non-BPS branes and/or
brane-antibrane systems. In this paper we hope to provide some
explanations and address various related issues at least in the
classical approximation by showing their relations to the
well-studied tubular D6 branes.

For this, we recall that the Kaluza-Klein dipole solution
\cite{Gross:1983hb,Sorkin:1983ns} when embedded in eleven
dimensions, was successfully interpreted as a static
D6-brane--antiD6-brane configuration of type IIA string theory,
suspended in an external magnetic field, in
\cite{Dowker:1995sg,Sen:1997pr}. We remark that usually the
solution in $d=10$ (obtained  by the dimensional reduction of an
eleven dimensional one), can suffer from two kinds of
singularities, the curvature singularity and the conical
singularity. Just like in BPS branes, the curvature singularities
here appear at the locations of D6-brane and antiD6-brane. In
other words they represent the brane sources and should not be a
concern to us. On the other hand the conical singularity appears
because of the improper introduction of a background magnetic flux
or fluxbrane to the system under consideration (the eleven
dimensional origin of this is due to an improper choice of a pair
of coordinates not entirely independent of each other) by which
the attractive forces due to gravitational and magnetic
interactions remain unbalanced with the repulsive force induced by
the magnetic flux between the brane and the anti-brane. In this
case the system becomes classically unstable as well as dynamical.
For this reason, the conical singularity will be our real concern
here. The conical singularity never appears if the reduction of
the eleven dimensional KK dipole (regular) solution is on an
appropriate Killing direction with a `good' pair of independent
coordinates\footnote{When such a pair of coordinates (one of them
is a compact coordinate $x^{11}$ and the the other is an angular
coordinate introduced later) are entirely independent of each
other and have their respective correct period $2\pi R$ and
$2\pi$, we call them a `good' pair, otherwise a `bad' pair. In
fact, any choice of non-trivial twist on  the angular coordinate
in eleven dimensions is equivalent to introducing a new background
magnetic flux or fluxbrane to the underlying ten dimensional
configuration under consideration, therefore showing the
non-uniqueness of reduction as well as the background flux as
mentioned in \cite{Dowker:1995gb}. }  and thereby introducing a
well-tuned background magnetic flux or fluxbrane in $d=10$. This
magnetic flux in turn provides the repulsive force necessary to
counterbalance the attractive force between the brane and the
anti-brane, giving a static configuration of such unstable
systems. When the parameter `$a$' of the KK dipole solution (this
is the Euclideanized rotation parameter of the $d=4$ Kerr black
hole solution) vanishes, the corresponding eleven dimensional
configuration is the coincident D6-antiD6 brane system lifted to
$d = 11$ and precisely for this case, there is a good choice of
coordinate system in eleven dimensions such that when viewed from
ten dimensions, the resulting solution in $d = 10$ is free from
any magnetic flux \cite{Sen:1997pr}. The zero magnetic field
solution turns out to be just the chargeless $p=6$ configuration
found in \cite{Lu:2004ms} when two of the three parameters
characterizing the solution take some specific values and
therefore Brax et. al.'s \cite{Brax:2000cf} interpretation of the
latter solution as the D6-antiD6 system works for this
case\footnote{The same was successfully demonstrated in
\cite{Liang:2001s} for D6-antiD6 system with a net D6-brane charge
using Kerr-NUT solution lifted to ten dimensions.}.

The situation is not so clear for the lower dimensional ($p<6$)
branes. On one side we have chargeless non-supersymmetric
$p$-brane solutions found in \cite{Lu:2004ms} and in order for the
lower dimensional branes to have the interpretation of
brane-antibrane system (as for $p=6$) one may think that a higher dimensional
dipole-like solution (with higher form gauge field) should exist
on the other side. However, no such solutions are known to exist
in the literature. If we recall how the Kaluza-Klein dipole
solution in $d=5$, which is crucial for its interpretation of
monopole-antimonopole (or D6-antiD6 system when embedded in
$d=10$) system was obtained, we find that it is obtained from the
Euclidean Kerr black hole solution in $d=4$ with an extra time
direction added (this can be embedded in $d=11$ by adding six more
flat spatial directions whose reduction on S$^1$ gives D6-antiD6
system). It might seem that lower dimensional brane-antibrane
systems can similarly be obtained from the higher dimensional
($d>4$) Euclidean rotating black holes \cite{Myers:1986un} (with
rotation in only one plane), but the solution one generates by
this procedure in $d=10$ are the D6-branes with tubular topology
R$^{1,p}$ $\times$ S$^{6-p}$ called the tubular D6-branes
\cite{Emparan:2001rp}. The $p=6$ case is the one discussed above
and $p=0$ corresponds to a D6-brane with spherical topology S$^6$.
As for the case of $p=6$, the tubular D6-brane solutions could
also suffer from conical singularities if the background magnetic
flux or from the eleven dimensional viewpoint  a twisted coordinate is
not properly chosen. However, as before, here also a well-tuned
background magnetic flux can be introduced
through a dimensional reduction  along an appropriate
Killing direction of the corresponding eleven dimensional regular
solution with a `good' choice of twisted coordinate.  The magnetic
flux then provides the necessary repulsive force to balance the
gravitational and electromagnetic forces to prevent the
contraction of the tubular D6-brane and gives rise to a static
configuration. When the Euclidean rotation parameter `$a$' (which
measures in an appropriate coordinate system the tubular deformation
of the metric for the tubular D6 brane with the brane topology
R$^{1,p}$ $\times$ S$^{6 - p}$ from the maximal spherical
symmetry S$^{8-p}$), is put to zero, the tubular brane deforms
back to restore the maximal spherical symmetry S$^{8 - p}$, giving
rise to the non-supersymmetric chargeless $p$-brane. For this case
again there is a `good' choice of coordinate system in $d = 11$
such that when viewed from ten dimensions, it makes the solution
free from any magnetic flux and the solution can be interpreted as
static unstable system such as brane-antibrane or non-BPS brane
systems. Interestingly, these are precisely the three parameter
non-supersymmetric $p$-brane solutions found in \cite{Lu:2004ms}
with two of the three parameters restricted to take some specific
values.

In this paper we proceed from backwards. Our primary goal is to
give a proper interpretation of the ten dimensional
non-supersymmetric $p$-brane solutions found in \cite{Lu:2004ms}
by making their connections to bubbles and tubular D6-branes. So,
we first look at the $d=10$ non-supersymmetric $p$-brane solution
in \cite{Lu:2004ms} and after making a coordinate transformation
(this amounts to going to the Schwarzschild coordinate) we find
that for some specific values of two of the three parameters
characterizing the solution, the resulting solution takes the form
of bubbles \cite{Witten:1981gj} when lifted to eleven dimensions.
This means that the $p$-brane solutions are regular in $d=11$,
with the periodicity of the eleventh direction properly chosen,
just like some BPS branes (for example, NS5-brane or D2-brane).
Next we like to understand whether these well-understood bubbles
can give physically meaningful ten dimensional solutions and
whether these static solutions can be interpreted as unstable
systems like brane-antibrane systems or non-BPS branes. For this
purpose we bring out their relations to the well-understood
tubular D6-branes. We find that the ten dimensional
non-supersymmetric $p$-brane solutions \cite{Lu:2004ms} for $p\leq
6$ (with two of the parameters taking specific values as
mentioned) is just a special limit of the tubular D6-brane
solution with topology R$^{1,p}$ $\times$ S$^{6-p}$, when the
Euclidean rotation parameter `$a$' vanishes and the choice of
coordinate system in $d = 11$ is such (with a `good' pair) that
when viewed from ten dimensions, the resulting configuration is
free from any magnetic flux. This connection provides further
evidence for the interpretation of Brax et. al. \cite{Brax:2000cf}
in type IIA theory and also indicates that the localized non-BPS
branes or brane-antibrane systems with lower brane dimensionality
can be obtained from tubular D6-branes when proper limit is taken.
With the exchange of non-BPS brane and brane-antibrane we can also
give same interpretation in type IIB theory in the low energy
supergravity set up.

Without further ado, let us begin with the magnetically charged
non-supersymmetric $p$-brane solutions (static) in ten space-time
dimensions \cite{Lu:2004ms}, \bea ds^2 &=& F^{\frac{p+1}{8}}
(H{\tilde {H}})^{\frac{2}{7-p}} \left(dr^2 + r^2
d\Omega_{8-p}^2\right) + F^{-\frac{7-p}{8}}\left(-dt^2 +
\sum_{i=1}^p (dx^i)^2 \right)\nn e^{2\phi} &=& F^{-a}
\left(\frac{H}{\tilde {H}}\right)^{2\delta},\qquad F_{[8-p]}\,\,\,
=\,\,\, b\,\,{\rm Vol}(\Omega_{8-p}) \eea where\footnote{The
parameter $\theta$ here should not be confused with the angular
coordinate introduced later. Also the parameter $a$ which is the
dilaton coupling here should not be confused with the rotation
parameter `$a$' introduced later.} \bea F(r) &=& \cosh^2\theta
\left(\frac{H}{\tilde H}\right)^{\alpha} - \sinh^2\theta
\left(\frac{\tilde H}{H}\right)^{\beta}\nn H(r) &=& 1+
\frac{\omega^{7-p}}{r^{7-p}}\nn \tilde {H}(r) &=& 1 -
\frac{\omega^{7-p}}{r^{7-p}} \eea In the above the metric is given
in the Einstein frame. $a$ is the dilaton coupling which take
values $a=(p-3)/2$ for D$p$-branes and $a=(3-p)/2$ for NSNS
branes. Also $\theta$, $\alpha$, $\beta$, $\delta$, $\omega$ are
integration constants and $b$ is the charge parameter. However,
all these parameters are not independent and they satisfy the
following relations, \bea
&&\alpha-\beta\,\,\, = \,\,\,a\delta\\
&&\frac{1}{2} \delta^2 + \frac{1}{2}\alpha(\alpha-a\delta)
\,\,\, =\,\,\,
\frac{8-p}{7-p}\\
&&b \,\,\,=\,\,\, (7-p) \omega^{7-p} (\alpha+\beta) \sinh2\theta
\eea It is clear from above that the non-supersymmetric $p$-brane
solutions are characterized by three independent parameters,
namely, $\alpha$, $\theta$ and $\omega$. We would like to remark
that in order to bring connections with the non-BPS branes (or
equal number of brane-antibrane system) as well as with tubular
branes, we need to set $F_{[8-p]}$ to zero and this can be
achieved by setting the parameter $b=\theta = 0$ (see eq.(5)).
This implies that the $p$-brane solutions are chargeless. Now
since in this case the function $F(r)$ appearing in eq.(2) takes
the form $F(r)=\left(H/{\tilde H}\right)^{\alpha}$, the solution
(1) therefore simplifies to \bea ds^2 &=& (H{\tilde
{H}})^{\frac{2}{7-p}} \left(\frac{H}{\tilde
H}\right)^{\frac{p+1}{8} \alpha}\left(dr^2 + r^2
d\Omega_{8-p}^2\right) + \left(\frac{H}{\tilde
H}\right)^{-\frac{7-p}{8}\alpha}\left(-dt^2 + \sum_{i=1}^p
(dx^i)^2 \right)\nn e^{2\phi} &=& \left(\frac{H}{\tilde
{H}}\right)^{-a\alpha + 2\delta},\qquad F_{[8-p]}\,\,\, =\,\,\, 0
\eea Note here that now the solutions are characterized by only
two parameters $\alpha$ and $\omega$. The parameters $\alpha$ and
$\delta$ are again related by eq.(4). We now make a coordinate
transformation \be r = \tilde{r}
\left(\frac{1+\sqrt{f}}{2}\right)^{\frac{2}{7-p}} \ee where we
have defined $f = 1 - 4\omega^{7-p}/\tilde{r}^{7-p}$. The
coordinate transformation (7) implies \be \tilde{r} = r \left(1+
\frac{\omega^{7-p}}{r^{7-p}}\right)^{\frac{2}{7-p}} = r
H^{\frac{2}{7-p}} \ee From the above two eqs.(7) and (8) we can
obtain, \bea H &=& 1 + \frac{\omega^{7-p}}{r^{7-p}} =
\frac{2}{\sqrt{f}+1}\nn \tilde{H} &=& 1 -
\frac{\omega^{7-p}}{r^{7-p}} = \frac{2\sqrt{f}}{\sqrt{f}+1} \eea
Substituting (9) in eq.(6), the chargeless non-supersymmetric
$p$-brane solutions take the form \bea ds^2 &=& f^{\frac{1}{7-p} -
\frac{p+1}{16} \alpha} \left(\frac{d\tilde{r}^2}{f} + \tilde{r}^2
d\Omega_{8-p}^2\right) + f^{\frac{7-p}{16}\alpha}\left(-dt^2 +
\sum_{i=1}^p (dx^i)^2 \right)\nn e^{2\phi} &=& f^{\frac{a\alpha -
2\delta}{2}},\qquad F_{[8-p]}\,\,\, =\,\,\, 0 \eea where the
function $f$ is as defined before. For now, we can limit ourselves
to consider the above chargeless  solutions in ten dimensional
type IIA theory\footnote{Once we have interpreted the solutions as
representing either D-brane-antiD-brane or non-BPS D brane in Type
IIA theory, then this approach works for IIB theory also since a
supergravity description of a D$p$-anti D$p$ (non-BPS D$p$) system
in IIA gives a non-BPS D$p$ (D$p$-antiD$p$) in IIB theory.}
therefore they can be uplifted to $d=11$. The corresponding
solution is given as, \bea ds_{11}^2 &=& f^{\frac{a\a-2\d}{3}}
(dx^{11})^2 + f^{-\frac{a\a-2\d}{24}+\frac{1}{7-p}-
\frac{p+1}{16}\a} \left(\frac{ d\tilde{r}^2}{f} + \tilde{r}^2
d\Omega_{8-p}^2\right)\nn & & \qquad\qquad\qquad\qquad\qquad +
f^{-\frac{a\a-2\d}{24}+\frac{7-p}{16}\a} \left(-dt^2 +
\sum_{i=1}^{p} (dx^i)^2 \right) \eea where we have used $ds_{11}^2
= e^{4\phi/3} (dx^{11})^2 + e^{-\phi/6} ds^2$, with $ds^2$ the ten
dimensional line element in the Einstein frame. It can now be
easily checked that if the parameters $\a$ and $\d$ take the
following values,
 \bea \a
&=& \frac{2}{7-p}\nn \d &=& \frac{a}{7-p} - \frac{3}{2} \eea then
the eleven dimensional metric takes a very simple form
as\footnote{Note that for D6-brane, $\a=2$ and $\d=0$ and in this
case the similarity between the non-supersymmetric D6-brane with
zero charge (which can also be interpreted as D6-antiD6-brane
system with zero net charge) and the GPS
\cite{Gross:1983hb,Sorkin:1983ns} dipole solution embedded in
eleven dimensions has been noticed in
refs.\cite{Brax:2000cf,Liang:2001s}.}, \be ds_{11}^2 = f
(dx^{11})^2 + \frac{d\tilde{r}^2}{f} + \tilde{r}^2 d\Omega_{8-p}^2
- dt^2 + \sum_{i=1}^p (dx^i)^2 \ee The above solutions (13) can be
recognized as the analytic continuation ($t \to i x^{11}$, $x^{11}
\to it$) of $(11-p)$-dimensional Schwarzschild solution with $p$
flat spatial directions added. The configuration (13) is defined
for $\tilde r \ge 4^{1/(7 - p)} \omega$ and describes a minimal
$(8-p)$-dimensional sphere with $p$ flat directions in space when
$\tilde r = 4^{1/(7 - p)} \omega$, i.e., tubular bubbles in
unstable equilibrium \cite{Witten:1981gj}. The eleven dimensional
solutions are perfectly regular with $x^{11}$ having the
periodicity $2\pi R = 4\pi 4^{1/(7-p)} \omega/(7-p)$. However, if
we reduce the solutions along $x^{11}$ to ten dimensions the
solutions take the following form, \bea ds^2 &=& f^{\frac{1}{8}}
\left(\frac{d\tilde{r}^2}{f} + \tilde{r}^2 d\Omega_{8-p}^2\right)
+ f^{\frac{1}{8}}\left(-dt^2 + \sum_{i=1}^p (dx^i)^2 \right)\nn
e^{2\phi} &=& f^{\frac{3}{2}}, \eea which, as expected, is just
(10) when the relations in (12) are employed.
 Clearly these solutions look singular
at $\tilde r= r_H = 4^{1/(7-p)} \omega$. The minimal tubular
surface R$^p$ $\times$ S$^{8-p}$, the fixed point set of the
Killing vector $\partial_{x^{11}}$, shrinks to zero size at
$\tilde r=r_H$ and therefore represents a $p$-brane-like null
singularity \cite{Dowker:1995sg}. But the singularity here is not
a concern as mentioned earlier since it is due to the brane source
and we know that the ten dimensional description breaks down there
just like in BPS case. If we indeed want to study it, we have to
lift the configuration to eleven dimensions where the
corresponding metric (13) is regular\footnote{Note that the metric
(13) can be extended to the region of $\tilde r >0$ with a naked
singularity at $\tilde r = 0$. This naked singularity happens in
many other cases and is understood due to the brane sources
located at $\tilde r = r_H$. For this reason, we can limit to
define the metric a valid one only for $\tilde r \ge r_H$ which is
also consistent with the reduced ten dimensional metric (14) where
$\tilde r$ can only be defined for $\tilde r \ge r_H$.}.

Now let us rewrite eq.(13) in terms of some twisted coordinate
(there are many such choices as pointed out in
\cite{Dowker:1995gb,Dowker:1995sg}) as follows. We first write the
line element of the unit $(8-p)$-dimensional sphere as \be
d\Omega_{8-p}^2 = d\theta^2 + \sin^2\theta d\varphi^2 +
\cos^2\theta d\Omega_{6-p}^2 \ee where $0 \leq \theta \leq \pi/2$,
$0 \leq \varphi \leq 2\pi$. Then we introduce a new coordinate by
\be \tilde{\varphi} = \varphi - \frac{x^{11}}{R} \ee where
$R=2\cdot 4^{1/(7-p)}\omega/(7-p)$ is as given before. Now if
instead of taking $x^{11}$ and $\varphi$ as the independent
coordinates in eq.(13) we take $x^{11}$ and $\tilde{\varphi}$ as
the independent coordinates and express eq.(13) in terms of these
latter coordinates then we find that $ds_{11}^2$ in (13) takes the
form\footnote{Henceforth we will drop `tilde' on $r$ in eqs.(13)
and (14) for brevity.}, \bea ds_{11}^2 &=& -dt^2 + \sum_{i=1}^p
(dx^i)^2 + \left(f+\frac{r^2\sin^2\theta}{R^2}\right)
\left(dx^{11} + \frac{r^2
\sin^2\theta}{R\left(f+\frac{r^2\sin^2\theta}{R^2}\right)}
d\tilde{\varphi} \right)^2\nn & & \qquad +
\frac{f}{\left(f+\frac{r^2\sin^2\theta}{R^2}\right)}
r^2\sin^2\theta d\tilde{\varphi}^2 + \frac{dr^2}{f} + r^2
d\theta^2 + r^2 \cos^2\theta d\Omega_{6-p}^2 \eea Reducing these
solutions to ten dimensions according to the prescription given
earlier we find the corresponding ten dimensional solutions to
have the form, \bea ds^2 &=&
\left(f+\frac{r^2\sin^2\theta}{R^2}\right)^{\frac{1}{8}}\left(-dt^2
+ \sum_{i=1}^p (dx^i)^2\right) +
\left(f+\frac{r^2\sin^2\theta}{R^2}\right)^{\frac{1}{8}}
\left(\frac{dr^2}{f} + r^2 d\theta^2 \right.\nn & & \left. + r^2
\cos^2\theta d\Omega_{6-p}^2 + \frac{f}{\left(f + \frac{r^2
\sin^2\theta}{R^2}\right)} r^2\sin^2\theta
d\tilde{\varphi}^2\right)\nn e^{\frac{4\phi}{3}} &=&
\left(f+\frac{r^2\sin^2\theta}{R^2}\right)\,\,\,=\,\,\, 1 -
\frac{4\omega^{7-p}}{r^{7-p}} + \frac{r^2 \sin^2\theta
(7-p)^2}{4\cdot 4^{2/(7-p)} \omega^2}\nn A_{\tilde{\varphi}} &=&
\frac{r^2\sin^2\theta e^{-\frac{4\phi}{3}}}{R} \eea Now it can be
easily checked that in these forms the solutions (18) are regular
at $r=r_H$ except at $\theta=0$. In fact one can check for any
possible conical singularities of these solutions away from
$\theta=0$ by looking at the periodicity of the angular coordinate
$\tilde{\varphi}$ and see whether it fails to satisfy the usual
periodicity of $2\pi$ or not. From the form of the metric (18) we
indeed find the circumference to radius ratio
\cite{Emparan:2001rp} \be
\left.\frac{2\pi}{\sqrt{g_{rr}}}\,\,\frac{d\sqrt
{g_{\tilde{\varphi}\tilde{\varphi}}}}{dr}
\right|_{r\to r_H} = 2\pi \ee

This result can be understood as follows. Recall
that the original configurations (13) is perfectly regular for $r\geq
r_H$ with well-defined completely independent `good' pair of
coordinates $\varphi$ and $x^{11}$ having respective periodicities
$2\pi$ and $2\pi R$. This implies that (13) is free from any conical singularity
and so is (14). The eleven dimensional configuration (17) which
is obtained from (13) by a change of coordinates (16),
another `good' pair of coordinates ($\tilde\varphi$ and $x^{11}$)
is also free from conical singularity. So, the reduction of (17)
along the corresponding Killing
direction should not produce any conical singularity in the
configuration (18). This is what is manifested in (19).

Another
way to understand (19) is that the reduced solutions (18) can be
regarded as describing a D6-brane with the tubular topology
R$^{1,p}$ $\times$ S$^{6-p}$ located at $r = r_H$ and $\theta =
0$, held in static equilibrium by the external magnetic flux (or
fluxbrane) $B=1/R$. In order to understand this we remark
following \cite{Dowker:1995sg,Emparan:2001rp} that, we have
essentially reduced the eleven dimensional solutions (13) to
$d=10$ by choosing independent coordinates $\tilde\varphi$ and
$x^{11}$ along the Killing vector $\partial_{x^{11}}$. The fixed
points of this Killing vector are at $r=r_H = 4^{1/(7-p)} \omega$
and $\theta=0$ which are located on the tubular surface R$^{1,p}$
$\times$ S$^{6-p}$. As discussed in \cite{Dowker:1995sg}, each of
these points on the sphere\footnote{Bear in mind that monopoles at
antipodal points on the sphere have opposite charges so the net
magnetic charge on the sphere is zero.} S$^{6 - p}$ carries a
Kaluza-Klein monopole charge and therefore the solutions (18)
represent tubular D6 branes with the topology just mentioned.
However, these configurations themselves in the zero background
magnetic field would be unstable and have either time dependence
or the presence of conical singularities because of the brane
tension of the tubular D6-brane. In the particular reduction from
(17) to (18) the attractive forces due to gravitational and
magnetic interactions are cancelled by the repulsive force,
induced by the external magnetic flux (or fluxbrane), which
manifests itself by the absence of conical singularity.  So again,
we understand the configuration (18) free from any conical singularity
from the ten dimensional viewpoint through the force balance.
We thus end
up with a static tubular D6 configuration in $d = 10$ with the
magnetic flux $B=1/R=(7-p)/(2r_H)$ which can be read from (16)
following \cite{Dowker:1995gb,Dowker:1995sg}.

The above discussion
clearly indicates that any choice of non-trivially twisted
coordinate such as (16) in eleven dimensions introduces a
background magnetic flux or fluxbrane in ten dimensions when the
reduction is done along the relevant Killing direction (here with
independent coordinates $\tilde\varphi$ and $x^{11}$ the Killing
direction is $\partial_{x^{11}}$). However, we must caution that
the value of the magnetic flux is unphysically high and as such
this makes the ten dimensional interpretation of the solutions
problematic since the magnetic flux produces curvature in the
non-compact direction which is of the same order as the inverse
size of the compact directions \cite{Dowker:1995gb}. So this
reduction to ten dimensions employed above is used only for the
purpose of showing the force balance and therefore the absence of
conical singularity in (18).

We would like to point out that the tubular D6 brane above has no
net magnetic charge but has a magnetic field due to the magnetic
charge distribution on the sphere S$^{6 - p}$, as given in (18).
This tubular D6 brane can naturally  be interpreted as a pair of
(D6-D$p$)-anti(D6-D$p$)-brane system with their respective brane
topologies R$^{1, p}$ $\times$ S$_+^{6 - p}$ and R$^{1, p}$
$\times$ S$_-^{6-p}$. Here S$^{6 - p}_+$ (S$_-^{6 - p}$)
represents a semi $(6 - p)$-sphere with S$^{6 - p}$ = S$_+^{6 -
p}$ $\bigcup$ S$_-^{6 - p}$.  The notation (D6-D$p$)
(anti(D6-D$p$)) means a D6 (anti D6)  with the above topology and
its own positive (negative) magnetic charge distributed on a
semi-sphere S$^{6 - p}_+$ (S$_-^{6 - p}$) and  with a D$p$ (anti
D$p$) positive (negative) electric-like charge uniformly
distributed on $R^{1,p}$. The D$p$ and anti D$p$ are coincident
and just like the previously discussed coincident D6-antiD6 case
their net charges are always vanishing in the supergravity
configuration in any coordinate system as no electric or
magnetic-like form field associated with the D$p$--anti D$p$ brane
appears. This picture is consistent with what has been discussed
when $p = 6$. We know that the tubular D6-brane appears by a
coordinate twist given in (16) or from the ten dimensional point
of view when a background magnetic flux $B = 1/R$ is introduced.
If we use the original good coordinates $\varphi$ and $x^{11}$,
neither the tubular D6-brane nor the background $B$ is present (as
is clear by the absence of $A_\varphi$ in (14)). This indicates
that the magnetic charges on the sphere S$^{6 - p}$ are caused by
the background magnetic flux through polarization. This further
implies that the tubular D6 brane is created by the background
magnetic flux, too. No background magnetic flux means no tubular
D6 brane. As will be discussed later in the paper, the natural
reduction from eleven to ten is the one without the twist on the
angular coordinate, i.e., from the ten dimensional point of view
without the introduction of the background magnetic flux,
therefore no tubular D6 brane. This indicates that the system
under consideration i.e. (14) indeed represents a $p$-brane-anti
$p$-brane or non-BPS $p$-brane system. The static nature of (14)
as we have shown to follow from the absence of conical singularity
(which can be understood through the force balance as discussed
above), can also be understood by the application of Birkhoff's
theorem to this configuration possessing the maximal spherical
symmetry. This fact was used for $p=6$ by Brax et. al.
\cite{Brax:2000cf} to argue that this configuration can represent
brane-antibrane or non-BPS brane in the supergravity set up.

In the following we will generalize the above discussion by
introducing a Euclidean rotation
parameter `$a$' to the configuration (13). By this, we can avoid unphysically high
magnetic field and also make the connection of (14), interpreted as representing
$p$-brane-anti $p$-brane or non-BPS $p$-brane system, with the tubular D6 branes in a
more general setup. This can be done, as mentioned
in the introductory remarks, by generalizing the KK
dipole solution of GPS \cite{Gross:1983hb, Sorkin:1983ns} with an
addition of a time direction to the Euclidean rotating black holes (with
rotation in only one plane) in higher than four dimensions
\cite{Myers:1986un} and then
embedding the resulting solution into eleven dimensions by adding some
flat spatial directions. The solutions can be written in the form,


\bea
ds_{11}^2 &=& -dt^2 + \sum_{i=1}^p (dx^i)^2 + \left(\frac{\Delta +
a^2 \sin^2\theta} {\Sigma}\right)\left(dx^{11} - \frac{M r^{p-5} a
\sin^2\theta}{\Delta + a^2 \sin^2\theta} d\varphi\right)^2\nn & &
\qquad\qquad + \left[\Sigma\left(\frac{dr^2}{\Delta} +
d\theta^2\right) + \frac{\Sigma \Delta \sin^2\theta}{\Delta + a^2
\sin^2\theta}d\varphi^2 + r^2 \cos^2\theta d\Omega_{6-p}^2\right]
\eea
where `$a$' is the Euclidean rotation parameter and `$M$' is the
Euclidean mass parameter. $\Delta$ and $\Sigma$ are defined as
$\Delta=r^2 - Mr^{p-5} -a^2$ and $\Sigma=r^2-a^2\cos^2\theta$.
For $p=6$ the solution (20) reduces exactly to the GPS dipole solution
embedded in eleven dimensions \cite{Sen:1997pr}. Note that when
the parameter $a\to 0$,
\be
\frac{\Delta + a^2 \sin^2\theta}{\Sigma} =
1- \frac{Mr^{p-5}}{r^2 - a^2\cos^2\theta} \rightarrow
1 -\frac{M}{r^{7-p}} \equiv 1 - \frac{r_H^{7-p}}{r^{7-p}} = f
\ee
and the above solution reduces to the solution (13). We will
discuss more about this reduction as we proceed.
These
solutions (20) have `Euclidean' horizons at $r=r_0$, where $r_0$ is the
greatest root of the equation $\Delta=0$ and is given by \be r_0^2
= M r_0^{p-5} + a^2 \ee Let us define  \be R= \frac{1}{\kappa} =
\frac{2 M r_0^{p-4}}{(7-p) r_0^2 - (5-p) a^2}, \,\,\, \Omega =
\frac{a r_0^{5-p}}{M} \ee with $\kappa$ the surface gravity and
$\Omega$ the Euclidean angular velocity, and \be \hat{\varphi} =
\varphi - \Omega x^{11}.\ee The solutions (20) are then regular
for $r\geq r_0$, provided $x^{11}$ has the periodicity $2\pi R$
and $\hat{\varphi}$ has the periodicity $2\pi$. This implies that
$\varphi$ has to be correlated with $x^{11}$ in having a
periodicity \be 2\pi \frac{\Omega}{\kappa} = 2\pi \Omega R =
\frac{4\pi a r_0}{(7-p) r_0^2 - (5-p) a^2} \ee in order to have
identifications of spacetime points.
 There are various ways one can reduce these solutions to
ten dimensions. It can be checked that if we choose independent
coordinates $\hat\varphi$ and $x^{11}$ and  reduce (20) along
the Killing vector $\partial_{x^{11}}$ (the Killing vector is
$\partial_{x^{11}} + \Omega \partial_\varphi$ if expressed in
terms of $\varphi$ and $x^{11}$ coordinates), then the fixed point
sets are the whole horizon $r=r_0$. Also, since the coefficient of
$(dx^{11})^2$ vanishes in this case, the corresponding ten
dimensional solutions appear singular. This is a (null) curvature
singularity, indicating the location of the brane source, not a
concern to us as discussed earlier. This case is very much like
what we had in reducing (13) to (14). There are no conical
singularities here, indicating that the repulsive force induced by
the background magnetic flux $B = \Omega$ on the tubular D6 brane
cancels the attractive forces due to the gravitational and
magnetic interactions. Another way to reduce the solutions (20) to
ten dimensions is to take $x^{11}$ and $\varphi$ as independent
coordinates and reduce (20) along the Killing vector
$\partial_{x^{11}}$ (Note that this Killing vector is not the same
as the above). We end up with \bea ds^2 &=& \left(\frac{\Delta +
a^2 \sin^2\theta}{\Sigma}\right)^{\frac{1}{8}} \left(-dt^2 +
\sum_{i=1}^p (dx^i)^2\right) + \left(\frac{\Delta + a^2
\sin^2\theta} {\Sigma}\right)^{\frac{1}{8}}
\left[\frac{\Sigma}{\Delta} dr^2 + \Sigma d\theta^2 \right.\nn & &
\left. + \frac{\Sigma \Delta \sin^2\theta}{\Delta + a^2
\sin^2\theta} d\varphi^2 + r^2 \cos^2 \theta
d\Omega_{6-p}^2\right]\nn e^{2\phi} &=& \left(\frac{\Delta + a^2
\sin^2\theta}{\Sigma}\right)^{\frac{3}{2}}\nn A_{\varphi} &=&
-\frac{M r^{p-5} a \sin^2\theta}{\Delta + a^2 \sin^2\theta} \eea
Note that the above solutions are the more general tubular D6-brane
branes than the solutions (18) involving an additional parameter `$a$'.
The solutions (26) match exactly with (14) for $a = 0$ with now
$A_{\varphi} = 0$ as expected. However, we expect a problem to
arise when $a \neq 0$ since $x^{11}$ and $\varphi$ cannot be taken
to be entirely independent of each other, a so-called `bad' choice
of coordinates. Their periodicities are actually correlated as
mentioned earlier. As a result of this, conical singularities
arise at $r = r_0$ for these solutions (26). The underlying
picture for these singularities from the ten dimensional view is
that the attractive forces due to the magnetic and gravitational
interactions on the tubular D6-brane are not balanced. There is
still another way to reduce the solutions (20) where the problem
does not arise and that is taking the following $\tilde\varphi$
and $x^{11}$ as independent coordinates and reducing along the
Killing vector $\partial_{x^{11}}$. Introducing the twisted
coordinate \be \tilde{\varphi} = \varphi - (\Omega - \kappa)x^{11}
\quad \Rightarrow \quad \varphi = \tilde{\varphi} +
(\Omega-\kappa)x^{11} \ee and treating $\tilde{\varphi}$ and
$x^{11}$ as the independent coordinates in (20) we find the
reduced solutions in $d=10$ as, \bea ds^2 &=&
\Lambda^{\frac{1}{8}}\left(-dt^2 + \sum_{i=1}^p (dx^i)^2\right) +
\Lambda^{\frac{1}{8}}\left[\Sigma\left(\frac{dr^2}{\Delta} +
d\theta^2\right) + \frac{\Delta \sin^2\theta}{\Lambda}
d\tilde{\varphi}^2 + r^2 \cos^2\theta d\Omega_{6-p}^2\right],\nn
e^{2\phi} &=& \Lambda^{\frac{3}{2}},\nn A_{\tilde{\varphi}} &=&
\frac{(\Omega-\kappa)(r^2-a^2)\sin^2\theta}{\Lambda} - \frac{M
r^{p-5} a \sin^2\theta}{\Sigma\Lambda} \left(1 -
a(\Omega-\kappa)\sin^2\theta\right), \eea where \be \Lambda = 1-
\frac{Mr^{p-5}}{\Sigma}\left(1 -
a(\Omega-\kappa)\sin^2\theta\right)^2 +
(\Omega-\kappa)^2(r^2-a^2)\sin^2\theta. \ee Now it can be checked
that with the above value of the background magnetic flux \be B =
\Omega - \kappa = \frac{(p-5)r_0 + (p-7)a} {2r_0(r_0+a)}, \ee
there are no conical singularities in the solutions (28). Indeed
we find from the metric in (28) that \be
\left.\frac{2\pi}{\sqrt{g_{rr}}}\,\,
\frac{d\sqrt{g_{\tilde{\varphi}
\tilde{\varphi}}}}{dr}\right|_{r\to r_0} = 2\pi. \ee Actually,
this fact can be anticipated from the outset. Recall that the
reduction is along $x^{11}$ but now taking $x^{11}$ and
$\tilde\varphi$ as independent coordinates. From (27) and (24), we
have $\tilde\varphi = \hat{\varphi} + \kappa x^{11}$. Note that
$\hat{\varphi}$, unlike $\varphi$, is independent of $x^{11}$ and
has its own period of $2\pi$ and therefore this relation is
similar to the case in (16). Since $\tilde\varphi$ can be taken
completely independent of $x^{11}$, a good choice of coordinates,
no conical singularities should arise and the total force
acting on the tubular D6-brane must be balanced.

It is clear from (30) that the value of the magnetic flux $B$ can
be made arbitrarily small by adjusting $\Omega$
\cite{Dowker:1995sg} and consequently the ten dimensional
solutions (28) can be trusted. This will also be true for the $B=\Omega$
case. So, the point we want to make is
that although the eleven dimensional solution (20) (or (13) which
is the $a\to 0$ limit of (20)) is perfectly regular and meaningful
in the sense described below (24), their ten dimensional reduction
is not always physically acceptable. So, for example, the reduced
solution (26) has problems of interpretation since it contains
conical singularity for general $a$. The reduced solution (28) is
acceptable only for $a\gg  M^{1/(7-p)}$ which gives
$|B| = |\Omega-\kappa| \ll \kappa$. The $d=10$ non-supersymmetric $p$-brane
solutions (14) are also physically acceptable reduction from (20)
in the $a\to 0$ limit. This can be understood in two ways as follows.
We can take the solution (20) directly and put $a=0$. This will give (13)
and then it could be reduced to ten dimensional solution (14) in the
usual way. Otherwise, we can choose $\hat\varphi$ and $x^{11}$ as
independent pair of `good' coordinates in (20) then reduce it along the
Killing vector $\partial_{x^{11}}$ and finally put $a=0$. This will
also give (14). In other words, various properties regarding solution
(14) can be studied in a similar fashion as the case discussed previously
with $a=0$. For $a$ of the same order as
$M^{1/(7-p)}$, only the eleven dimensional description (20)
or its other
forms related to this one by a change of the so-called `good'
coordinates is appropriate as pointed out in \cite{Sen:1997pr}.
This implies that the best description in making various
connections among the chargeless non-supersymmetric $p$-brane,
bubbles and tubular D6 branes is the eleven dimensional one.

To summarize, in this paper we have tried to interpret the non-supersymmetric,
three parameter $p$-brane solutions \cite{Lu:2004ms} of type II
string theories as $p$-brane-anti $p$-brane or non-BPS $p$-brane systems.
We have also tried to clarify their static nature and the various
singularities by bringing out their relations to tubular bubbles and
tubular D6-branes. First we have shown that when the charge parameter
is put to zero and one other parameter takes a specific value then
the resulting non-supersymmetric $p$-brane solution in different coordinate
system takes the form of tubular bubbles and is regular when uplifted
to $d=11$ given in eq.(13). We argued that the corresponding ten dimensional
solution (14) can be interpreted as $p$-brane-anti $p$-brane or non-BPS $p$
brane systems. Since it is known that these systems are unstable one may
be sceptical how they could be given by static configurations like (14).
We have argued at least in the classical supergravity set up, that there
is no net force acting on such systems conforming the view that unstable
systems in question can be represented by static configurations. To show
this we brought out the connections of the solutions (14) with tubular
D6-branes.

We have seen that in eleven dimensions and in the absence of
rotation parameter `$a$' there are various ways one can choose a
pair of coordinates (one is the compact coordinate $x^{11}$ and
the other is an angular coordinate $\varphi$) and with each choice
(twist) a nonvanishing magnetic flux is introduced in the reduced
(along a related Killing direction) ten dimensional solution (a
tubular D6-brane). In other words one can say that tubular
D6-brane is created by the magnetic flux through polarizations. We
emphasized that when our choice is `good', then in ten dimensions
we get a force balance for a specific value of the magnetic field
(which may not always be physically acceptable) on the tubular
D6-brane otherwise not. In this case the solution also does not
suffer from conical singularity. We have generalized the above
discussion by introducing a rotation parameter `$a$' to the
configuration under consideration. Here we understood the picture
of the absence of conical singularity, force balance and the role
of the various choices of coordinates more clearly. So, for
example, in this case there are various choices of the angular
coordinates such as $\varphi$ in (20), $\hat\varphi$ in (24) or
$\tilde\varphi$ in (27), but not all the choices are `good'. One
such `bad' choice of coordinates is $\varphi$ in (20). As we have
pointed out, for this choice, the periodicities of the coordinates
$\varphi$ and $x^{11}$ are correlated and not really independent
of each other as in the case of other choices like $\hat\varphi$
or $\tilde\varphi$. When we reduce these solutions to ten
dimensions, only for `good' choice of coordinates we get a force
balance and don't have any conical singularity. Two such solutions
are (18) (when there is no parameter `$a$') and (28) (in fact (18)
is the $a \to 0$ limit of (28) with $R$ replaced by $-R$). The
absence of conical singularity for the tubular D6-brane (17) must
imply the same and the corresponding force balance for its ten
dimensional version (18). This in turn implies the same for the
tubular bubbles (13), therefore its ten dimensional version (14)
since (13) is related to (17) by a simple `good' coordinate
change. The manifest absence of conical singularity in either (13)
or (14) is consistent with the above. This very fact explains the
static nature of the tubular bubble (13) or the $p$-brane--anti
$p$-brane (or non-BPS $p$ brane) solution (14) and is in
accordance with Birkhoff's theorem that the spherically symmetric
solution of vacuum Einstein's equation must be static. This also
supports the view that the maximally spherically symmetric
solution (14) which is static can indeed represent unstable
systems like $p$-brane-anti$p$-brane or non-BPS $p$-brane systems
(note that the solutions (14) have the required symmetry and also
they are chargeless as it should be).

It is therefore clear that at the classical supergravity
level there is no force acting between a $p$-brane and
an anti$p$-brane when they are coincident and the system is given by
a static configuration (as is also the case for non-BPS
$p$-brane). This implies that the interaction which causes the
brane annihilation or tachyon condensation occurs only at the
quantum level. Precisely for this reason, we can have a
supergravity configuration appearing as a static configuration at
each moment of the brane annihilation with its ADM mass related to
the interaction energy of the system labelled by the off-shell
tachyon as a parameter at that moment. So, we can use a continuous
family of supergravity configurations with its parameters relating
to the tachyon parameter to study the brane annihilation process.
For example, we can set the ADM mass calculated from the gravity
configuration equal to twice the brane tension plus the tachyon
potential as we did in \cite{Lu:2004dp,Bai:2005jr}. Finally we
would like to mention that even though in this paper we have
interpreted only a subclass of the non-supersymmetric $p$-brane
solutions obtained in \cite{Lu:2004ms}, it seems to us that a similar
interpretation can be given for the general chargeless solution
(10) or (11) as well since it appears that (15)-(19) can be similarly carried
through and they are just sufficient for this purpose. However, whether
the general solution (1) with non-zero $p$-brane charge can have similar
interpretation remains to be seen.

\vspace{.5cm}

\noindent {\bf Acknowledgements}

\vspace{2pt}

We would like to thank Sudipta Mukherji for helpful discussions.
JXL acknowledges support by grants from the Chinese Academy of
Sciences and grants from the NSF of China with Grant No: 90303002,
10588503 and 10535060.


\begin{thebibliography}{99}

\bibitem{Horowitz:1991cd}
  G.~T.~Horowitz and A.~Strominger,
  ``Black strings and P-branes,''
  Nucl.\ Phys.\ B {\bf 360}, 197 (1991).

\bibitem{Duff:1994an}
  M.~J.~Duff, R.~R.~Khuri and J.~X.~Lu,
  ``String solitons,''
  Phys.\ Rept.\  {\bf 259}, 213 (1995)
  [arXiv:hep-th/9412184].

\bibitem{Sen:1999mg}
  A.~Sen,
  ``Non-BPS states and branes in string theory,''
  arXiv:hep-th/9904207.

\bibitem{Aharony:1999ti}
  O.~Aharony, S.~S.~Gubser, J.~M.~Maldacena, H.~Ooguri and Y.~Oz,
  ``Large N field theories, string theory and gravity,''
  Phys.\ Rept.\  {\bf 323}, 183 (2000)
  [arXiv:hep-th/9905111].

\bibitem{Strominger:1996sh}
  A.~Strominger and C.~Vafa,
  ``Microscopic Origin of the Bekenstein-Hawking Entropy,''
  Phys.\ Lett.\ B {\bf 379}, 99 (1996)
  [arXiv:hep-th/9601029].

\bibitem{Maldacena:1997re}
  J.~M.~Maldacena,
  ``The large N limit of superconformal field theories and supergravity,''
  Adv.\ Theor.\ Math.\ Phys.\  {\bf 2}, 231 (1998)
  [Int.\ J.\ Theor.\ Phys.\  {\bf 38}, 1113 (1999)]
  [arXiv:hep-th/9711200].

\bibitem{Witten:1998qj}
  E.~Witten,
  ``Anti-de Sitter space and holography,''
  Adv.\ Theor.\ Math.\ Phys.\  {\bf 2}, 253 (1998)
  [arXiv:hep-th/9802150].

\bibitem{Gubser:1998bc}
  S.~S.~Gubser, I.~R.~Klebanov and A.~M.~Polyakov,
  ``Gauge theory correlators from non-critical string theory,''
  Phys.\ Lett.\ B {\bf 428}, 105 (1998)
  [arXiv:hep-th/9802109].

\bibitem{Sen:1999nx}
  A.~Sen and B.~Zwiebach,
  ``Tachyon condensation in string field theory,''
  JHEP {\bf 0003}, 002 (2000)
  [arXiv:hep-th/9912249].

\bibitem{Berkovits:2000hf}
  N.~Berkovits, A.~Sen and B.~Zwiebach,
  ``Tachyon condensation in superstring field theory,''
  Nucl.\ Phys.\ B {\bf 587}, 147 (2000)
  [arXiv:hep-th/0002211].

\bibitem{Brax:2000cf}
  P.~Brax, G.~Mandal and Y.~Oz,
  ``Supergravity description of non-BPS branes,''
  Phys.\ Rev.\ D {\bf 63}, 064008 (2001)
  [arXiv:hep-th/0005242].

\bibitem{Lu:2004dp}
  J.~X.~Lu and S.~Roy,
  ``Supergravity approach to tachyon condensation on the brane-antibrane
  system,''
  Phys.\ Lett.\ B {\bf 599}, 313 (2004)
  [arXiv:hep-th/0403147].


\bibitem{Zhou:1999nm}
  B.~Zhou and C.~J.~Zhu,
  ``The complete black brane solutions in D-dimensional coupled gravity
  system,''
  arXiv:hep-th/9905146.

\bibitem{Lu:2004ms}
  J.~X.~Lu and S.~Roy,
  ``Static, non-SUSY p-branes in diverse dimensions,''
  JHEP {\bf 0502}, 001 (2005)
  [arXiv:hep-th/0408242].

\bibitem{Bai:2005jr}
  H.~Bai, J.~X.~Lu and S.~Roy,
  ``Tachyon condensation on the intersecting brane-antibrane system,''
  JHEP {\bf 0508}, 068 (2005)
  [arXiv:hep-th/0506115].

\bibitem{Kobayashi:2004ay}
  S.~Kobayashi, T.~Asakawa and S.~Matsuura,
  ``Open string tachyon in supergravity solution,''
  Mod.\ Phys.\ Lett.\ A {\bf 20}, 1119 (2005)
  [arXiv:hep-th/0409044].

\bibitem{Lu:2004xi}
  J.~X.~Lu and S.~Roy,
  ``Delocalized, non-SUSY p-branes, tachyon condensation and tachyon matter,''
  JHEP {\bf 0411}, 008 (2004)
  [arXiv:hep-th/0409019].

\bibitem{Lu:2005ju}
  J.~X.~Lu and S.~Roy,
  ``Non-SUSY p-branes delocalized in two directions, tachyon condensation and
  T-duality,''
  JHEP {\bf 0506}, 026 (2005)
  [arXiv:hep-th/0503007].

\bibitem{Lu:2005jc}
  J.~X.~Lu and S.~Roy,
  ``Fundamental strings and NS5-branes from unstable D-branes in
  supergravity,''
  arXiv:hep-th/0508045.

\bibitem{Gross:1983hb}
  D.~J.~Gross and M.~J.~Perry,
  ``Magnetic Monopoles In Kaluza-Klein Theories,''
  Nucl.\ Phys.\ B {\bf 226}, 29 (1983).

\bibitem{Sorkin:1983ns}
  R.~D.~Sorkin,
  ``Kaluza-Klein Monopole,''
  Phys.\ Rev.\ Lett.\  {\bf 51} (1983) 87.

\bibitem{Dowker:1995sg}
  F.~Dowker, J.~P.~Gauntlett, G.~W.~Gibbons and G.~T.~Horowitz,
  ``Nucleation of $P$-Branes and Fundamental Strings,''
  Phys.\ Rev.\ D {\bf 53}, 7115 (1996)
  [arXiv:hep-th/9512154].

\bibitem{Sen:1997pr}
  A.~Sen,
  ``Strong coupling dynamics of branes from M-theory,''
  JHEP {\bf 9710}, 002 (1997)
  [arXiv:hep-th/9708002].

\bibitem{Dowker:1995gb}
  F.~Dowker, J.~P.~Gauntlett, G.~W.~Gibbons and G.~T.~Horowitz,
  ``The Decay of magnetic fields in Kaluza-Klein theory,''
  Phys.\ Rev.\ D {\bf 52}, 6929 (1995)
  [arXiv:hep-th/9507143].

\bibitem{Liang:2001s}
  Y.~C.~Liang and E.~Teo,
  ``Black diholes with unbalanced magnetic charges,''
  Phys.\ Rev.\ D {\bf 64}, 024019 (2001)
  [arXiv:hep-th/0101221].

\bibitem{Emparan:2001rp}
  R.~Emparan,
  ``Tubular branes in fluxbranes,''
  Nucl.\ Phys.\ B {\bf 610}, 169 (2001)
  [arXiv:hep-th/0105062].

\bibitem{Myers:1986un}
  R.~C.~Myers and M.~J.~Perry,
  ``Black Holes In Higher Dimensional Space-Times,''
  Annals Phys.\  {\bf 172}, 304 (1986).

\bibitem{Witten:1981gj}
  E.~Witten,
  ``Instability Of The Kaluza-Klein Vacuum,''
  Nucl.\ Phys.\ B {\bf 195}, 481 (1982).


\end{thebibliography}
\end{document}